\documentclass[aps,prd,reprint,amsmath,amssymb,groupedaddress,nofootinbib,floatfix,longbibliography]{revtex4-2}
\usepackage[T1]{fontenc}
\usepackage{lmodern}
\usepackage{bm}
\usepackage{graphicx}
\usepackage{mathrsfs}
\usepackage{hyperref}
\usepackage{capt-of}
\hypersetup{colorlinks=true,citecolor=blue,linkcolor=blue,urlcolor=blue}
\newcommand{\dd}{\mathrm{d}}
\newcommand{\ii}{\mathrm{i}}

\newcommand{\wt}{\mathcal{K}_F}

\newcommand{\vev}[1]{\langle #1\rangle}
\newcommand{\RR}{\mathbb{R}}
\begin{document}

\title{Quantum field theory with scale-dependent weighted measures:\\
nonlinear momentum representations, vacuum fluctuations, and causal structure}
\author{S. Gonz\'alez-Salud}
\email{santiago.g.salud@gmail.com}
\affiliation{Facultad de Ciencias F\'isico-Matem\'aticas, Benem\'erita Universidad Aut\'onoma de Puebla, 72570 Puebla, Mexico}
\author{R. Cartas-Fuentevilla}
\email{rcartas@ifuap.buap.mx}
\affiliation{Instituto de F\'isica ``Luis Rivera Terrazas'', Benem\'erita Universidad Aut\'onoma de Puebla, AP J-48, 72570 Puebla, Mexico}
\date{October 2026}
\begin{abstract}
We reconsider the construction of free quantum fields through momentum-space weighted measures, introducing an independent energy scale that restores the conventional theory at low momenta. The resulting family is characterized by a dimensionless spectral function and includes, as a distinguished case, the spatial weight associated with the Euclidean Snyder algebra. We discuss its relation to modified dispersion relations and nonlinear momentum coordinates, and introduce an observer-covariant spectral weight defined using a timelike four-velocity. For the complex Klein--Gordon field we obtain the equal-time algebra, conserved weighted observables, vacuum fluctuations, and two-point functions in arbitrary spatial dimension. The ultraviolet conditions for their finiteness are established, and the nonlocal causal structure is analyzed in terms of a spatial convolution of the ordinary Pauli--Jordan distribution. An illustrative interacting extension and numerical profiles of the kernels and amplitudes are also presented.
\end{abstract}
\maketitle

\section{Introduction}\label{sec:intro}
It is well known that ultraviolet divergences arise in quantum field theory (QFT), requiring regularization and renormalization procedures in order to define physical observables. While these methods provide a highly successful and systematic framework, it remains worthwhile to investigate whether the short-distance singular structure of QFT can be softened directly at the level of its spectral formulation.

Among the possible departures from the standard framework, one may modify the canonical algebra, the dispersion relation, or the spectral measure defining the quantum fields. The first possibility naturally appears in theories of noncommutative coordinates and generalized uncertainty principles \cite{Snyder1947,Kempf1995}; the second has been extensively considered in approaches with modified relativistic kinematics, including doubly special relativity (DSR) \cite{MagueijoSmolin2002}. A third possibility, which is the one pursued here, is to reconsider the momentum-space measure itself \cite{Cartas2022,Cartas2025}.

Recently we have initiated a program for constructing finite field observables by considering Lorentz-symmetry-breaking weighted measures \cite{Cartas2022}. The results, although incipient, provide us a glimpse of the program's promise. First, the vacuum expectation value of the energy associated with the weighted field can be finite without normal ordering; second, the equal-time commutators can be represented by finite spatial kernels; and third, one obtains families of propagators whose light-cone singular behavior is softened \cite{Cartas2025}. These results suggest that the standard Dirac delta distribution is not the only possible object entering the equal-time quantum field algebra.

Whereas these observations may be of interest, an important difficulty remains in the previous constant-$s$ deformation: the same exponent acts over the entire momentum range and no independent transition scale ensures the recovery of the ordinary theory in the infrared. Furthermore, the power of the frequency carries nonstandard mass dimensions, which makes the physical interpretation of a continuously varying exponent less immediate. The present work is intended to address these difficulties.

A useful clue comes from a different line of investigation. In \emph{The measure matters}, Amelino-Camelia \emph{et al.}\ showed that a class of modified dispersion relations (MDRs) can be represented, after a nonlinear change of momentum coordinates, by the conventional dispersion relation together with a modified integration measure \cite{MeasureMatters}. Related ideas also arise in the nonlinear action of the Lorentz group on momentum space, which lies at the heart of DSR \cite{MagueijoSmolin2002}. This correspondence gives a possible kinematical interpretation of the measure, but requires some care: an arbitrary momentum weight does not, by itself, imply an invariant energy scale or a deformed Lorentz symmetry.

Here we introduce a positive, dimensionless function $F_\Lambda$ with an independent crossover scale $\Lambda$, and study the replacement
\begin{equation}\label{eq:measure-intro}
    \frac{\dd^Dk}{2\omega_k}\rightarrow
    \dd\mu_F(\mathbf{k})=\frac{\dd^Dk}{2\omega_k}\,F_\Lambda(k),
\end{equation}
with $\omega_k=\sqrt{k^2+m^2}$. We shall see that a quadratic member of this family reproduces the Hilbert-space measure arising from the Euclidean Snyder algebra \cite{LuStern2012}. We further introduce an observer-covariant extension of the weight, which reduces to the original momentum-dependent function in the rest frame of the observer. The same family allows one to control the regularity of the field, the expectation values in the vacuum, and the departure from exact microcausality through a common scale.

This work is organized as follows. In Sec.~\ref{sec:formalism} we introduce the spectral family, examine its Snyder and MDR motivations, and present an observer-covariant extension. In Sec.~\ref{sec:klein-gordon} we consider the complex Klein--Gordon field, its conserved observables, and vacuum expectation values. Section~\ref{sec:causality} is devoted to the two-point functions, light-cone behavior, and spacelike tails. An illustrative interacting extension is described in Sec.~\ref{sec:applications}; Sec.~\ref{sec:conclusions} collects our conclusions and perspectives. Technical derivations are presented in the Appendices.

\section{Spectral measures and their physical interpretation}\label{sec:formalism}

\subsection{A scale-dependent family}\label{sec:family}
The most immediate requirements on the weight are positivity, rotational invariance, and recovery of the usual infrared limit. Accordingly, we impose
\begin{align}
    &F_\Lambda(k)>0,\qquad F_\Lambda(0)=1,\nonumber\\
    &F_\Lambda(k)\longrightarrow1\quad(k/\Lambda\longrightarrow0),\\
    &F_\Lambda(k)\sim(\Lambda/k)^r\quad(k/\Lambda\longrightarrow\infty),\nonumber
\end{align}
with $r>0$. A simple family satisfying these conditions is
\begin{equation}\label{eq:weight}
    F_\Lambda(k;n,r)=\left[1+\left(\frac{k}{\Lambda}\right)^n\right]^{-r/n},
    \qquad n,r>0.
\end{equation}
In contrast with a constant power of $\omega_k$, this function is dimensionless, while the scale $\Lambda$ determines when the spectral suppression becomes relevant. In particular, the ordinary theory is recovered for $\Lambda\to\infty$ at fixed momentum.

Motivated by our previous $s$ weighted-deformation \cite{Cartas2022,Cartas2025}, as well as by logarithmic scaling exponents appearing in studies of dimensional reduction \cite{MeasureMatters,DimReduction}, we define an effective weighted power,
\begin{equation}\label{eq:seff}
    s_{\rm eff}(k)=1-\frac{\dd\ln F_\Lambda(k)}{\dd\ln k}
    =1+r\frac{(k/\Lambda)^n}{1+(k/\Lambda)^n}.
\end{equation}
It interpolates monotonically between $s_{\rm eff}(0)=1$, $s_{\rm eff}(\Lambda)=1+r/2$, and $s_{\rm UV}=1+r$. This definition is related to the running exponents used to characterize momentum-space dimensional flow, but should not be identified directly with the spectral dimension of spacetime.

For comparison, an exponential family $F_\Lambda^{\exp}(k)=\exp[-(k/\Lambda)^n]$ would give $s_{\rm eff}=1+n(k/\Lambda)^n$. In what follows we shall restrict ourselves to Eq.~\eqref{eq:weight}, with parti\-cular attention to $n=2$ and $r>D+1$.

\subsection{A measure from the Snyder algebra}\label{sec:snyder}
There is an independent and particularly useful motivation for selecting $n=2$. Let us consider the Euclidean Snyder algebra in $D$ spatial dimensions \cite{Snyder1947,LuStern2012},
\begin{align}
    [\hat x_i,\hat k_j]&=\ii\left(\delta_{ij}
    +\frac{\hat k_i\hat k_j}{\Lambda^2}\right),\nonumber\\
    [\hat x_i,\hat x_j]&=\frac{\ii}{\Lambda^2}
    (\hat x_i\hat k_j-\hat x_j\hat k_i),\nonumber\\
    [\hat k_i,\hat k_j]&=0.
    \label{eq:snyder-algebra}
\end{align}
On wave functions $f(\mathbf k)$, the coordinate operator admits the representation
\begin{equation}\label{eq:snyder-op}
    \hat x_i=\ii\left(\delta_{ij}+\frac{k_i k_j}{\Lambda^2}\right)
    \partial_{k_j}.
\end{equation}
Requiring this operator to be symmetric with respect to the momentum-space scalar product $\int\dd^Dk\,w(k)f^*g$ fixes, up to an overall constant,
\begin{equation}\label{eq:snyder-measure}
    w(k)=\left(1+\frac{k^2}{\Lambda^2}\right)^{-(D+1)/2},
\end{equation}
as shown in Appendix~\ref{app:snyder}. Consequently, the Euclidean Snyder measure belongs exactly to the family \eqref{eq:weight}, with
\begin{equation}\label{eq:snyder-r}
    n=2,\qquad r=D+1,\qquad s_{\rm UV}=D+2.
\end{equation}
In $D=3$ this is the weight $(1+k^2/\Lambda^2)^{-2}$ employed in Ref.~\cite{LuStern2012}.

As we shall see, $r=D+1$ is also the borderline case for the convergence of the weighted zero-point energy. We therefore consider $r>D+1$ as a UV-finite extension of the Snyder-motivated family, retaining the quadratic transition while allowing the ultraviolet power to vary. The Snyder commutation relations admit symmetric rep\-resentations for arbitrary $r$, as discussed in Appendix~\ref{app:snyder}.

\subsection{Modified dispersion relations and nonlinear momenta}\label{sec:dsr}
The relation with modified dispersion relation (MDR) theories can be established through a change of momentum coordinates. For an isotropic weight, define a new radial coordinate $\rho(k)$ by
\begin{equation}\label{eq:rho-map}
    \rho^D(k)=D\int_0^k\dd q\,q^{D-1}F_\Lambda(q),
    \qquad
    \boldsymbol\kappa=\frac{\rho(k)}{k}\mathbf k.
\end{equation}
For $r>D$ the integral can be expressed in terms of the incomplete beta function,
\begin{equation}
\begin{split}
    \rho^D(k)&=\frac{D\Lambda^D}{n}
    B_{v(k)}\left(\frac Dn,\frac{r-D}{n}\right),\\
    v(k)&=\frac{(k/\Lambda)^n}{1+(k/\Lambda)^n}.
\end{split}
\end{equation}
Since $F_\Lambda(k)>0$, the transformation is monotone and satisfies
\begin{equation}\label{eq:jacobian}
    \dd^D\kappa=F_\Lambda(k)\,\dd^Dk.
\end{equation}
Thus, the weighted spatial measure can be expressed u\-sing the unweighted coordinate $\boldsymbol\kappa$, but with the modified mass-shell relation
\begin{equation}\label{eq:mdr-dual}
    E^2-k^2(\boldsymbol\kappa)=m^2,
    \qquad k(\boldsymbol\kappa)=\rho^{-1}(|\boldsymbol\kappa|).
\end{equation}
This is the dual description emphasized in Ref.~\cite{MeasureMatters}: a change of momentum coordinates can transfer the deformation between the integration measure and the dispersion relation.

For $r>D$, the new momentum coordinate has a finite range,
\begin{equation}\label{eq:rhomax}
    \kappa_{\max}^D=\frac{D\Lambda^D}{n}
    B\left(\frac Dn,\frac{r-D}{n}\right).
\end{equation}
This does not impose an upper bound on the energy, since $k(\boldsymbol\kappa)\to\infty$ as $\kappa\to\kappa_{\max}$. The corresponding nonlinear Lorentz transformations may be obtained by conjugating ordinary boosts with the inverse coordinate map,
\begin{equation}\label{eq:nonlinear-boosts}
    W_{\mathsf L}=\mathcal U^{-1}\circ\mathsf L\circ\mathcal U,
    \qquad
    \mathcal U(E,\boldsymbol\kappa)
    =(E,\mathbf k(\boldsymbol\kappa)).
\end{equation}
These transformations preserve the modified mass shell and the bounded momentum domain. Their existence, however, is not sufficient to establish a physically distinct doubly special relativity (DSR) theory: the transformation of the integration measure, field amplitudes, and interacting momenta must also be specified \cite{MagueijoSmolin2002}. We shall leave that for future work.

\subsection{An observer-covariant spectral measure}\label{sec:covariant}
The preceding relation suggests a more direct physical interpretation of our weight without modifying the Klein--Gordon dispersion relation. Let $u^\mu$ denote the constant, future-directed four-velocity of an observer, normalized by $u_\mu u^\mu=1$. The spatial momentum measured in its rest frame has magnitude
\begin{equation}\label{eq:observer-q}
    q^2(k,u)=(u_\mu k^\mu)^2-k_\mu k^\mu.
\end{equation}
We can therefore extend Eq.~\eqref{eq:weight} to the Lorentz-scalar function
\begin{equation}\label{eq:covariant-weight}
    F_\Lambda(k,u;n,r)=
    \left[1+\left(\frac{q(k,u)}{\Lambda}\right)^n\right]^{-r/n}.
\end{equation}
In the observer's rest frame, $u^\mu=(1,\mathbf0)$, we have $q=k$ and recover the original family exactly. In particular, for $n=2$,
\begin{equation}\label{eq:covariant-weight-n2}
    F_\Lambda(k,u;2,r)=
    \left[1+\frac{(u\cdot k)^2-k_\mu k^\mu}{\Lambda^2}\right]^{-r/2}.
\end{equation}
For a massless particle, $k_\mu k^\mu=0$ and $u\cdot k>0$ on the future light cone, so that
\begin{equation}\label{eq:covariant-massless}
    F_\Lambda(k,u;n,r)=
    \left[1+\left(\frac{u\cdot k}{\Lambda}\right)^n\right]^{-r/n}.
\end{equation}
Thus, the massless weight depends directly on the energy measured by the observer, whereas for massive particles it depends on the corresponding spatial momentum. Both prescriptions possess the same infrared limit and ultraviolet suppression in the observer's rest frame.

The corresponding on-shell measure is
\begin{equation}\label{eq:covariant-measure}
    \dd\mu_{F,u}(k)=\dd^{D+1}k\,
    \theta(k^0)\delta(k_\mu k^\mu-m^2)F_\Lambda(k,u).
\end{equation}
It is invariant under simultaneous proper orthochronous Lorentz transformations of $k^\mu$ and $u^\mu$, and reduces to $F_\Lambda(k)\dd^Dk/(2\omega_k)$ in the observer's rest frame. This is an observer-covariant family of measures, not an invariant scalar measure under boosts of $k^\mu$ with $u^\mu$ held fixed. The latter would require the weight to be constant on the ordinary massive mass shell. If $u^\mu$ is a fixed universal background, this distinction is the familiar one between observer covariance and particle Lorentz invariance \cite{ColladayKostelecky1998}. If instead $u^\mu$ characterizes the measuring apparatus, it labels the observer-dependent resolution of the field, and the construction is closely motivated by MDR/DSR ideas.

The same interpretation applies to the filtered field. For $n=2$ one may define
\begin{equation}\label{eq:covariant-filter}
    \phi_{F,u}=\mathcal T_{\Lambda,u}\phi,
    \quad
    \mathcal T_{\Lambda,u}=
    \left[1+\frac{\Box-(u\cdot\partial)^2}{\Lambda^2}\right]^{-r/4},
\end{equation}
with the operator understood as a Fourier multiplier on hypersurfaces orthogonal to $u^\mu$. In the rest frame this reduces to $\mathcal T_{\Lambda,u}=(1-\nabla^2/\Lambda^2)^{-r/4}$. The field is therefore resolved spatially with respect to the chosen observer. Its correlation functions form a covariant family under simultaneous transformations of their spacetime arguments and $u^\mu$; their nonzero spacelike tails are not eliminated by this covariance.

In what follows we shall restrict ourselves to a static inertial observer, $u^\mu=(1,\mathbf0)$, for simplicity.

\subsection{The weighted Fourier map and spatial resolution}\label{sec:fourier}
In the rest frame of the chosen observer, let $\mathscr F_\mu$ denote the Fourier transform from $L^2(\RR^D,\dd\mu_F)$ into $L^2(\RR^D,\dd^Dx)$,
\begin{equation}\label{eq:unitary-FT}
    (\mathscr F_\mu f)(\mathbf x)
    =\frac1{(2\pi)^{D/2}}
    \int\dd^Dk\,\sqrt{\frac{F_\Lambda(k)}{2\omega_k}}\,
    \mathrm e^{\ii\mathbf k\cdot\mathbf x}f(\mathbf k).
\end{equation}
With the corresponding inverse multiplication by $\sqrt{2\omega_k/F_\Lambda(k)}$, Eq.~\eqref{eq:unitary-FT} is unitary on its natural Hilbert spaces. It also suggests considering the weighted field as a finite-resolution observable associated with the ordinary auxiliary Klein--Gordon field,
\begin{equation}\label{eq:filter}
\begin{split}
    \widetilde\phi_F(t,\mathbf k)&=\sqrt{F_\Lambda(k)}\,
    \widetilde\phi(t,\mathbf k),\\
    \phi_F(t,\mathbf x)&=\int\dd^Dy\,W_F(\mathbf x-\mathbf y)
    \phi(t,\mathbf y),
\end{split}
\end{equation}
where $W_F=\mathcal F^{-1}[\sqrt{F_\Lambda}]$, with the Fourier normalization chosen so that the convolution contains no additional factors. The field $\phi$ may be viewed as an auxiliary perfectly resolved field, whereas $\phi_F$ represents its spatially filtered observable. The operator in Eq.~\eqref{eq:covariant-filter} extends this interpretation to an arbitrary inertial observer.

Of particular importance is the kernel
\begin{equation}\label{eq:kernel-def}
\begin{split}
    \wt(\mathbf X)&=\int\frac{\dd^Dk}{(2\pi)^D}
    F_\Lambda(k)\mathrm e^{\ii\mathbf k\cdot\mathbf X},\\
    \int&\dd^DX\,\wt(\mathbf X)=1.
\end{split}
\end{equation}
For $0<n\le2$ the functions \eqref{eq:weight} are positive definite, and $\wt$ is nonnegative by the Bochner--Schoenberg characterization \cite{Schoenberg1938}. For $n>2$ a nontrivial nonnegative normalized kernel is impossible, since the second derivative of $F_\Lambda$ at zero vanishes while the function is not constant. This provides a further reason for selecting $n=2$: it gives a genuine spatial averaging kernel.

For $n=2$, with $X=|\mathbf X|$, one obtains
\begin{equation}\label{eq:kernel-bessel}
    \wt(X)=\frac{2^{1-r/2}\Lambda^D}
    {(2\pi)^{D/2}\Gamma(r/2)}
    (\Lambda X)^{(r-D)/2}K_{(r-D)/2}(\Lambda X),
\end{equation}
where $K_\nu$ is the modified Bessel function. For $r>D$ its coincidence limit is
\begin{equation}\label{eq:kernel-zero}
    \wt(0)=\frac{\Lambda^D}{(4\pi)^{D/2}}
    \frac{\Gamma[(r-D)/2]}{\Gamma(r/2)}.
\end{equation}
In particular, the Snyder member $r=D+1$ reduces to
\begin{equation}\label{eq:snyder-kernel}
    \wt^{\rm Snyder}(X)=
    \frac{\Lambda^D}{2^D\pi^{(D-1)/2}\Gamma[(D+1)/2]}
    \mathrm e^{-\Lambda X}.
\end{equation}
It has a characteristic length of order $\Lambda^{-1}$. For $n=2$ the spatial root-mean-square width is
\begin{equation}\label{eq:rms}
    \ell_{\rm RMS}=\sqrt{\langle X^2\rangle}
    =\frac{\sqrt{rD}}{\Lambda}.
\end{equation}
The short-distance regularity depends on the ultraviolet power: finite $r$ generally gives only finitely many continuous derivatives, while $r>D+q$ is a sufficient condition for $\wt\in C^q$.

\begin{figure}[t]
\centering
    \includegraphics[
        width=\columnwidth,
        trim=1.5cm 0cm 0.5cm 0cm,
        clip
    ]{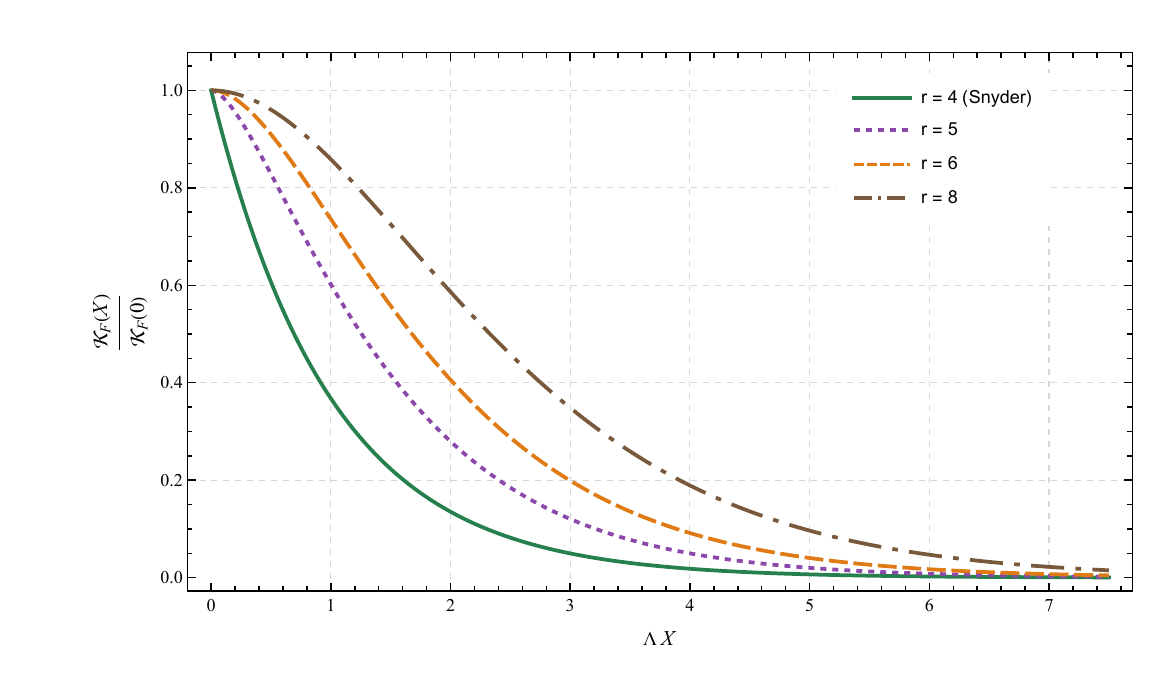}
\caption{Normalized spatial kernel for $n=2$, $D=3$ and different values of $r$. The Snyder value $r=4$ is included for comparison; ultraviolet finiteness of the weighted energy requires $r>4$.}\label{fig:kernel}
\end{figure}

\section{The complex Klein--Gordon field}\label{sec:klein-gordon}
\subsection{Free-field algebra and an effective action}\label{sec:action}
We now consider a complex Klein--Gordon field in $D+1$ dimensions, using natural units $c=\hbar=1$ and the metric $\eta_{\mu\nu}=\mathrm{diag}(+,-,\ldots,-)$. In the particle--antiparticle symmetric sector its mode expansion is
\begin{equation}\label{eq:field-expansion}
\begin{split}
    \hat\phi_F(t,\mathbf x)=\int&\frac{\dd^Dk}{(2\pi)^{D/2}}
    \sqrt{\frac{F_\Lambda(k)}{2\omega_k}}\\
    &\times\left[\hat a_{\mathbf k}\mathrm e^{-\ii\omega_k t+\ii\mathbf k\cdot\mathbf x}
    +\hat b_{\mathbf k}^{\dagger}\mathrm e^{\ii\omega_k t-\ii\mathbf k\cdot\mathbf x}\right],
\end{split}
\end{equation}
with $[a_{\mathbf k},a_{\mathbf p}^{\dagger}]=[b_{\mathbf k},b_{\mathbf p}^{\dagger}]=\delta^{(D)}(\mathbf k-\mathbf p)$ and all other oscillator commutators vanishing. The field satisfies $(\Box+m^2)\phi_F=0$ and has the usual dispersion relation. Writing $\pi_F=\dot\phi_F^\dagger$, the equal-time algebra takes the form
\begin{align}\label{eq:commutators}
    [\phi_F(t,\mathbf x),\pi_F(t,\mathbf y)]&=\ii\wt(\mathbf x-\mathbf y), \\
    [\phi_F(t,\mathbf x),\phi_F^\dagger(t,\mathbf y)]&=0, \\
    [\dot\phi_F(t,\mathbf x),\dot\phi_F^\dagger(t,\mathbf y)]&=0.
\end{align}
The ordinary equal-time canonical relation is recovered as $\Lambda\to\infty$ in the distributional sense.

The connection \eqref{eq:filter} also supplies a useful quadratic action. For the independent filtered field variable $\psi=\phi_F$, it reads
\begin{equation}\label{eq:filtered-action}
\begin{split}
    S_0[\psi]=\int\dd t\int&\frac{\dd^Dk}{(2\pi)^D}
    \frac1{F_\Lambda(k)}\\
    &\times\left[|\partial_t\widetilde\psi(t,\mathbf k)|^2
    -\omega_k^2|\widetilde\psi(t,\mathbf k)|^2\right].
\end{split}
\end{equation}
Equivalently, the free kinetic operator is the spatially nonlocal operator $F_\Lambda(\sqrt{-\nabla^2})^{-1}(\Box+m^2)$. Its equation of motion remains the usual Klein--Gordon equation; its canonical momentum, however, is $\Pi_\psi=F_\Lambda(\sqrt{-\nabla^2})^{-1}\dot\psi^\dagger$. Therefore $[\psi,\Pi_\psi]=\ii\delta^{(D)}$, while $[\psi,\dot\psi^\dagger]=\ii\wt$. In the chosen observer's rest frame, the action is local in time and nonlocal in space. With $u^\mu$ held fixed it is not invariant under ordinary boosts. Its observer-covariant extension, however, is invariant under simultaneous Lorentz transformations of the field and $u^\mu$, as made explicit in Sec.~\ref{sec:applications}.

\subsection{Energy, momentum and charge}\label{sec:observables}
Following the construction of the weighted observables in Ref.~\cite{Cartas2022}, let us form the standard quadratic densities out of $\phi_F$,
\begin{align}\label{eq:H-weighted}
    H_F&=\int\dd^Dx\,
    \left(|\dot\phi_F|^2+|\nabla\phi_F|^2+m^2|\phi_F|^2\right),\\
    \mathbf P_F&=-\int\dd^Dx\,
    \left(\dot\phi_F^\dagger\nabla\phi_F+
    \dot\phi_F\nabla\phi_F^\dagger\right),\label{eq:P-weighted}\\
    Q_F&=\ii\int\dd^Dx\,
    \left(\phi_F^\dagger\dot\phi_F-
    \dot\phi_F^\dagger\phi_F\right).
\label{eq:Q-weighted}
\end{align}
They are conserved in the free theory. Upon inserting Eq.~\eqref{eq:field-expansion}, one obtains
\begin{align}\label{eq:observables-modes}
    H_F&=\int\dd^Dk\,\omega_k F_\Lambda(k)
    \left(a_{\mathbf k}^\dagger a_{\mathbf k}
    +b_{\mathbf k}^\dagger b_{\mathbf k}+\delta_{\mathbf{x}}^{(D)}(0)\right),\\
    \mathbf P_F&=\int\dd^Dk\,\mathbf kF_\Lambda(k)
    \left(a_{\mathbf k}^\dagger a_{\mathbf k}
    +b_{\mathbf k}^\dagger b_{\mathbf k}\right), \\
    Q_F&=\int\dd^Dk\,F_\Lambda(k)
    \left(a_{\mathbf k}^\dagger a_{\mathbf k}
    -b_{\mathbf k}^\dagger b_{\mathbf k}\right).
\end{align}
The spatial integral of the vacuum momentum vanishes by parity. In Eq.~\eqref{eq:observables-modes}, $\delta_{\mathbf{x}}^{(D)}(0)=V/(2\pi)^D$ is the usual formal box-volume factor.

A distinction becomes important at this point. The conserved quantities \eqref{eq:observables-modes} are \emph{filtered, weighted observables}; they do not generate the ordinary time translations or global $U(1)$ transformations of the field with the canonical momentum derived from \eqref{eq:filtered-action}. The genuine Noether generators of that quadratic action contain the inverse weight in their coordinate-space definitions and become, in oscillator form, the usual unweighted sums $H_{\rm can}=\int\dd^Dk\,\omega_k(a^\dagger a+b^\dagger b+\delta^{(D)}(0))$ and $Q_{\rm can}=\int\dd^Dk(a^\dagger a-b^\dagger b)$. Accordingly, the finiteness of $\vev{H_F}$ established below is a statement about the energy carried by the filtered observable, not a removal of the formal vacuum constant from the canonical generator. This distinction is necessary for a consistent action-based interpretation.

\subsection{Vacuum fluctuations and zero-point energy}\label{sec:vacuum}
For the translation-invariant Fock vacuum $a_{\mathbf k}|0\rangle=b_{\mathbf k}|0\rangle=0$, the coincident two-point function is
\begin{equation}\label{eq:VEV-general}
    \vev{\phi_F(x)\phi_F^\dagger(x)}=
    \frac{\Omega_{D-1}}{2(2\pi)^D}
    \int_0^\infty\dd k\,\frac{k^{D-1}F_\Lambda(k)}{\omega_k},
\end{equation}
where $\Omega_{D-1}=2\pi^{D/2}/\Gamma(D/2)$, and the weighted vacuum-energy density is
\begin{equation}\label{eq:rho-general}
    \rho_{0,F}=\frac{\vev{H_F}}V=
    \frac{\Omega_{D-1}}{(2\pi)^D}
    \int_0^\infty\dd k\,k^{D-1}\omega_kF_\Lambda(k).
\end{equation}
The factor of two between the familiar real-scalar zero-point contribution and Eq.~\eqref{eq:rho-general} comes from the independent particle and antiparticle sectors. For the massless case ($D>1$), the integrals can be evaluated for arbitrary $n$:
\begin{align}\label{eq:VEV-zero-m}
    \vev{\phi_F\phi_F^\dagger}_{m=0}&=
    \frac{\Omega_{D-1}\Lambda^{D-1}}{2n(2\pi)^D}
    B\left(\frac{D-1}n,\frac{r-D+1}n\right),\\
    \rho_{0,F}^{m=0}&=
    \frac{\Omega_{D-1}\Lambda^{D+1}}{n(2\pi)^D}
    B\left(\frac{D+1}n,\frac{r-D-1}n\right).\label{eq:rho-zero-m}
\end{align}
For $n=2$ the massive expressions reduce to hypergeometric functions (Appendix~\ref{app:integrals}). The ultraviolet conditions are summarized in Table~\ref{tab:convergence}. In particular, the condition ensuring a finite vacuum energy and well-defined equal-time differentiated correlators is
\begin{equation}\label{eq:finite-r}
    \quad r>D+1\quad\qquad
    (s_{\mathrm{UV}}>D+2).
\end{equation}
This explains why the Snyder weight $r=D+1$ is especially interesting: it has a finite spatial commutator kernel and vacuum field fluctuation, but the energy \eqref{eq:rho-general} diverges logarithmically. An arbitrarily small positive increase of $r$ removes this logarithmic divergence.

\begin{table}[t]
\caption{Sufficient and, for the indicated positive integrands, necessary ultraviolet convergence conditions at fixed $m\ge0$. The $m=0$ coincident fluctuation additionally requires $D>1$ for infrared convergence.}\label{tab:convergence}
\begin{ruledtabular}
\begin{tabular}{ll}
Quantity & Ultraviolet requirement\\
$\wt(0)$, $\int\dd^Dk\,F_\Lambda$ & $r>D$\\
$\vev{\phi_F\phi_F^\dagger}$ & $r>D-1$\\
$\rho_{0,F}$ & $r>D+1$\\
$\int\dd^Dk\, k^q F_\Lambda$ & $r>D+q$
\end{tabular}
\end{ruledtabular}
\end{table}
\begin{figure}[t]
    \centering
    \includegraphics[
        width=\columnwidth,
        trim=2cm 0cm 0.5cm 0cm,
        clip
    ]{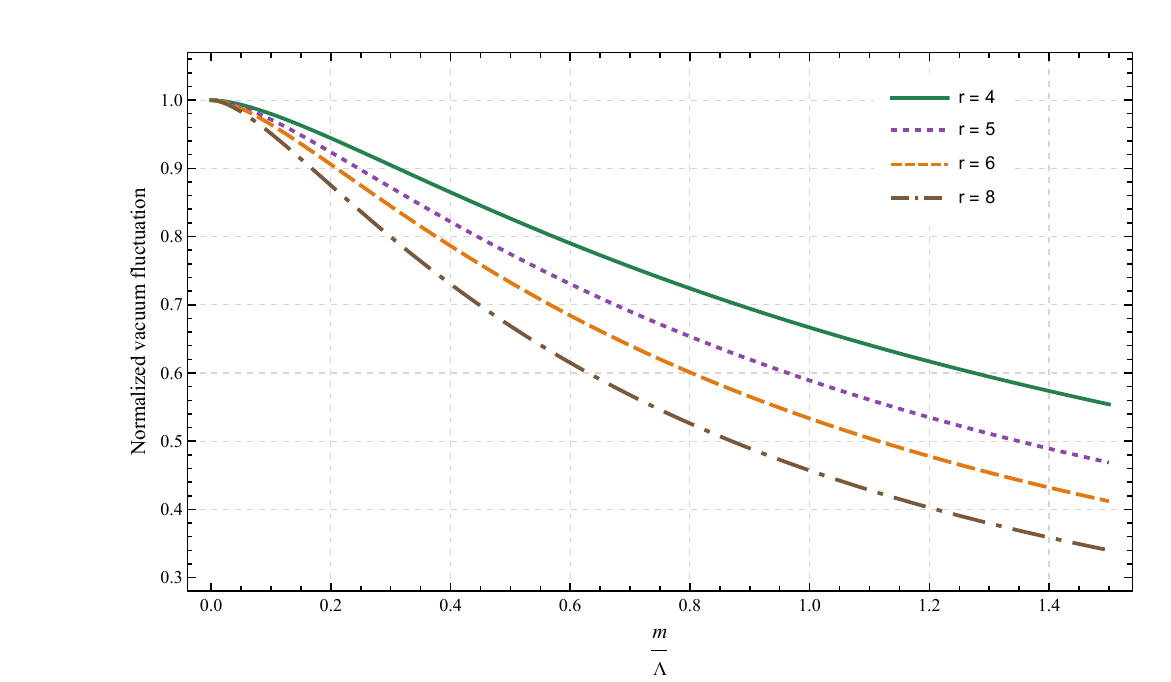}
    \caption{Coincident vacuum fluctuations for $n=2$, $D=3$,
    normalized by their massless values, as functions of
    $m/\Lambda$. All curves shown have finite field variance;
    the weighted vacuum energy is also finite when $r>4$.}
    \label{fig:vev}
\end{figure}

\subsection{Particle--antiparticle asymmetry}\label{sec:asymmetry}
The symmetric sector need not exhaust the available algebraic possibilities. Generalizing Eq.~\eqref{eq:field-expansion}, one may assign the two frequency sectors independent nonnegative factors $F_\pm(k)=F_\Lambda(k)[1\pm\eta(k)]$, where $\eta(k)$ is real, rotationally invariant, and $|\eta(k)|\le1$. The infrared condition $\eta(0)=0$ restores particle--antiparticle symmetry at small momenta. The resulting nontrivial equal-time commutators become
\begin{align}\label{eq:asymm-comm}
    [\phi_F,\dot\phi_F^\dagger]_{t=t'}&=\ii\wt(\mathbf X),\\
    [\phi_F,\phi_F^\dagger]_{t=t'}&=\mathcal K_{\rm NC}^{\phi}(\mathbf X),\\
    [\dot\phi_F,\dot\phi_F^\dagger]_{t=t'}&=\mathcal K_{\rm NC}^{\pi}(\mathbf X),
\end{align}
with
\begin{align}\label{eq:NC-kernels}
    \mathcal K_{\rm NC}^{\phi}(\mathbf X)&=
    \int\frac{\dd^Dk}{(2\pi)^D}\frac{F_\Lambda\eta}{\omega_k}
    \mathrm e^{\ii\mathbf k\cdot\mathbf X},\\
    \mathcal K_{\rm NC}^{\pi}(\mathbf X)&=
    \int\frac{\dd^Dk}{(2\pi)^D}F_\Lambda\eta\omega_k
    \mathrm e^{\ii\mathbf k\cdot\mathbf X}.
\end{align}
Thus $F_\Lambda$ determines the overall spatial resolution, while $\eta$ controls an independent asymmetry between the positive- and negative-frequency sectors. The quadratic action \eqref{eq:filtered-action} describes $\eta=0$; a consistent nonzero-$\eta$ dynamics requires a corresponding modification of the symplectic structure and will not be assumed here. The separate expressions are given in Appendix~\ref{app:algebra}.

\section{Two-point functions and causal structure}\label{sec:causality}
\subsection{Propagators and convolution representation}\label{sec:propagators}
Let us first restrict attention to the symmetric sector ($\eta=0$), thereby isolating the effects associated exclusively with the spectral deformation. Denoting $T=t-t'$ and $\mathbf X=\mathbf x-\mathbf x'$, the Wightman functions are
\begin{align}\label{eq:wightman}
    \Delta_+^F(T,\mathbf X)&=\int\frac{\dd^Dk}{(2\pi)^D}
    \frac{F_\Lambda(k)}{2\omega_k}
    \mathrm e^{-\ii\omega_kT+\ii\mathbf k\cdot\mathbf X},\\
    \Delta_-^F(T,\mathbf X)&=\Delta_+^F(-T,-\mathbf X).\nonumber
\end{align}
The corresponding Feynman propagator is
\begin{equation}\label{eq:G-momentum}
G_F(E,\mathbf k)=\frac{\ii F_\Lambda(k)}{E^2-k^2-m^2+\ii\epsilon},
\end{equation}
where the poles remain at the usual Klein--Gordon energies. In coordinate space it is the spatial convolution
\begin{equation}\label{eq:prop-conv}
G_F(T,\mathbf X)=\int\dd^Dy\,\wt(\mathbf y)
G_0(T,\mathbf X-\mathbf y).
\end{equation}
This identity makes the physical meaning of the construction particularly transparent: the weighted propagator is obtained by averaging the standard propagator over a finite spatial resolution. It also explains why the singular structure on the light cone can be softened without deforming the free dispersion relation.

The commutator, or Pauli--Jordan two-point function, is
\begin{equation}\label{eq:pj-general}
\begin{split}
    \Delta_C^F(T,\mathbf X)&=\vev{[\phi_F(x),\phi_F^\dagger(x')]}\\
    &=-\ii\int\frac{\dd^Dk}{(2\pi)^D}
    F_\Lambda(k)\frac{\sin(\omega_kT)}{\omega_k}
    \mathrm e^{\ii\mathbf k\cdot\mathbf X}.
\end{split}
\end{equation}
Using spherical symmetry it reduces to
\begin{equation}\label{eq:pj-radial}
\begin{split}
    \Delta_C^F(T,X)=-\ii\frac{X^{1-D/2}}{(2\pi)^{D/2}}
    &\int_0^\infty\dd k\,\frac{k^{D/2}F_\Lambda(k)}{\omega_k}\\
    &\times J_{D/2-1}(kX)\sin(\omega_kT).
\end{split}
\end{equation}
Equations~\eqref{eq:pj-general} and \eqref{eq:prop-conv} imply the exact convolution
\begin{equation}\label{eq:pj-convolution}
    \Delta_C^F(T,\mathbf X)=\int\dd^Dy\,\wt(\mathbf y)
    \Delta_C^0(T,\mathbf X-\mathbf y).
\end{equation}

Although the commutator vanishes at equal times, this does not establish microcausality, which requires a vanishing commutator for all spacelike separations. In fact, for $r>D$ we can expand Eq.~\eqref{eq:pj-general} around $T=0$ and obtain
\begin{equation}\label{eq:early-time}
\Delta_C^F(T,X)=-\ii T\wt(X)+o(T).
\end{equation}
For $n=2$, $\wt(X)>0$ at every finite $X>0$, so the commutator is generally nonzero at spacelike separation. It is therefore more appropriate to speak of \emph{suppressed spacelike correlations} than exact microcausality. 

\subsection{An exact spacelike representation}\label{sec:outside}
For $n=r=2$ an instructive closed representation is available in arbitrary spatial dimension. Introduce the Yukawa kernel
\begin{align}\label{eq:Yukawa}
    Y_D(X;\Lambda)&=\int\frac{\dd^Dk}{(2\pi)^D}
    \frac{\mathrm e^{\ii\mathbf k\cdot\mathbf X}}{k^2+\Lambda^2}\\
    &=\frac1{(2\pi)^{D/2}}
    \left(\frac\Lambda X\right)^{D/2-1}K_{D/2-1}(\Lambda X).
    \nonumber
\end{align}
and the odd function $S_\Lambda(T)$ satisfying
\begin{equation}\label{eq:S-def}
    S_\Lambda''+(m^2-\Lambda^2)S_\Lambda=0,
    \qquad S_\Lambda(0)=0,\quad S_\Lambda'(0)=1.
\end{equation}
Explicitly, 
\begin{equation}
    S_\Lambda(T) =
    \begin{cases}
        \displaystyle
        \frac{\sinh\left(T\sqrt{\Lambda^2-m^2}\right)}
             {\sqrt{\Lambda^2-m^2}},
        & \text{if } m < \Lambda, \\[8pt]
        T,
        & \text{if } m = \Lambda, \\[8pt]
        \displaystyle
        \frac{\sin\left(T\sqrt{m^2-\Lambda^2}\right)}
             {\sqrt{m^2-\Lambda^2}},
        & \text{if } m > \Lambda.
    \end{cases}
\end{equation}
With $G_D^{\rm ret}$ denoting the conventional retarded Klein--Gordon Green function, Eq.~\eqref{eq:pj-general} becomes
\begin{widetext}
\begin{equation}\label{eq:pj-resolvent}
    \Delta_C^{(2,2)}(T,X)=-\ii\Lambda^2\left[
    S_\Lambda(T)Y_D(X;\Lambda)
    -\operatorname{sgn}(T)\int_0^{|T|}\dd s\,S_\Lambda(s)
    G_D^{\rm ret}(|T|-s,X)\right].
\end{equation}
\end{widetext}
Since $G_D^{\rm ret}(t,X)=0$ for $X>|T|\ge0$, the integral vanishes throughout the spacelike region and
\begin{equation}\label{eq:pj-outside-r2}
    \Delta_C^{(2,2)}(T,X>|T|)=-\ii\Lambda^2 S_\Lambda(T)Y_D(X;\Lambda),
\end{equation}
Its large-$X$ behavior, with $T$ fixed, is therefore
\begin{equation}\label{eq:pj-asymp-r2}
    \Delta_C^{(2,2)}\sim
    -\frac{\ii\Lambda S_\Lambda(T)}{2(2\pi)^{(D-1)/2}}
    \left(\frac\Lambda X\right)^{(D-1)/2}
    \mathrm e^{-\Lambda X}.
\end{equation}
The $r=2$ case is an analytic building block; it does not satisfy the weighted-energy finiteness condition in $D=3$.

More generally, for the even powers $r=2q$ ($q=1,2,\ldots$), the same result gives an exact expression valid for arbitrary $D$ and $m$. Define $\Delta_C^{\mathrm{res}}$ as the Pauli--Jordan integral with the factor $(k^2+\Lambda^2)^{-1}$, \emph{without} the numerator $\Lambda^2$. Then
\begin{equation}\label{eq:general-even-spacelike}
    \Delta_C^{(2,2q)}(T,X)=
    \frac{\Lambda^{2q}}{(q-1)!}
    (-\partial_{\Lambda^2})^{q-1}
    \Delta_C^{\mathrm{res}}(T,X),
\end{equation}
where in the spacelike region
\begin{equation}\label{eq:res-exact}
    \Delta_C^{\mathrm{res}}(T,X)=-\ii S_\Lambda(T)Y_D(X;\Lambda).
\end{equation}
This differentiating formula yields the massive spacelike tail explicitly for every even $r$. In particular, $r=6$ gives a UV-finite weighted vacuum energy in $D=3$. For large spatial separation at fixed $T$, each derivative produces only algebraic factors multiplying the Yukawa exponential. The amplitude is consequently exponentially suppressed on the scale $\Lambda^{-1}$; near a moving light cone, $T\simeq X$, the more relevant distance is the spacelike gap $X-|T|$.

The behavior inside the light cone is more involved. For massive fields, the Klein--Gordon commutator contains an additional timelike contribution, whose oscillatory character can modify the relative amplitudes. This becomes particularly transparent by taking $T=\beta X$, with $\beta>1$ fixed. The radial integral then has a stationary point at $k_*=m/\sqrt{\beta^2-1}$, corresponding to modes whose group velocity satisfies $k_*/\omega_{k_*} =X/T$. In $D=3$ its leading contribution behaves as
\begin{equation}\label{eq:massive-timelike-asymptotic}
    \Delta_C^F(\beta X,X)\propto
    \frac{F_\Lambda(k_*)}{X^{3/2}}
    \cos\left(mX\sqrt{\beta^2-1}+\frac{\pi}{4}\right),
\end{equation}
up to a constant prefactor depending on $m$ and $\beta$. By comparison, for $n=2$, even $r$, and $0<m<\Lambda$, the on-cone and spacelike amplitudes are exponentially suppressed, with a decay factor $\exp[-\gamma(\beta)X]$, where $\gamma(\beta)=\Lambda-\beta\sqrt{\Lambda^2-m^2}$ for $0<\beta\leq1$. Thus, along the trajectories $T=(1\pm a)X$, with $0<a<1$, the exterior amplitude decreases faster than the on-cone one, whereas the timelike contribution has an oscillatory algebraic envelope. This explains why a hierarchy may be recovered asymptotically between the envelopes, although the commutator itself does not satisfy a universal pointwise ordering.

The preceding behavior is illustrated in Fig.~\ref{fig:pj-massive-comparison}. At smaller separations, the relative amplitudes depend on the spacetime region, while at larger distances the timelike contribution develops the oscillatory behavior described above. In particular, its algebraic envelope can be distinguished from the exponentially suppressed on-cone and spacelike contributions.

\onecolumngrid

\noindent
\begin{minipage}{\textwidth}
\centering
\includegraphics[
    width=0.49\textwidth,
    trim=1cm 0cm 0.5cm 0cm,
    clip
]{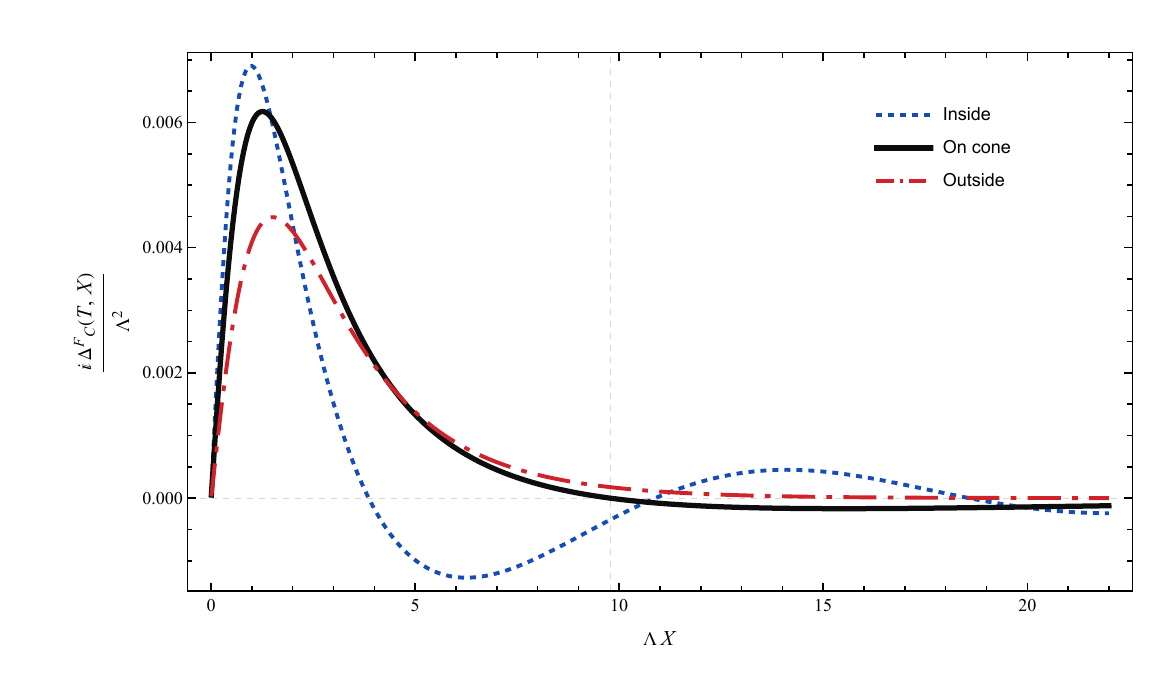}
\hfill
\includegraphics[
    width=0.49\textwidth,
    trim=1cm 0cm 0.5cm 0cm,
    clip
]{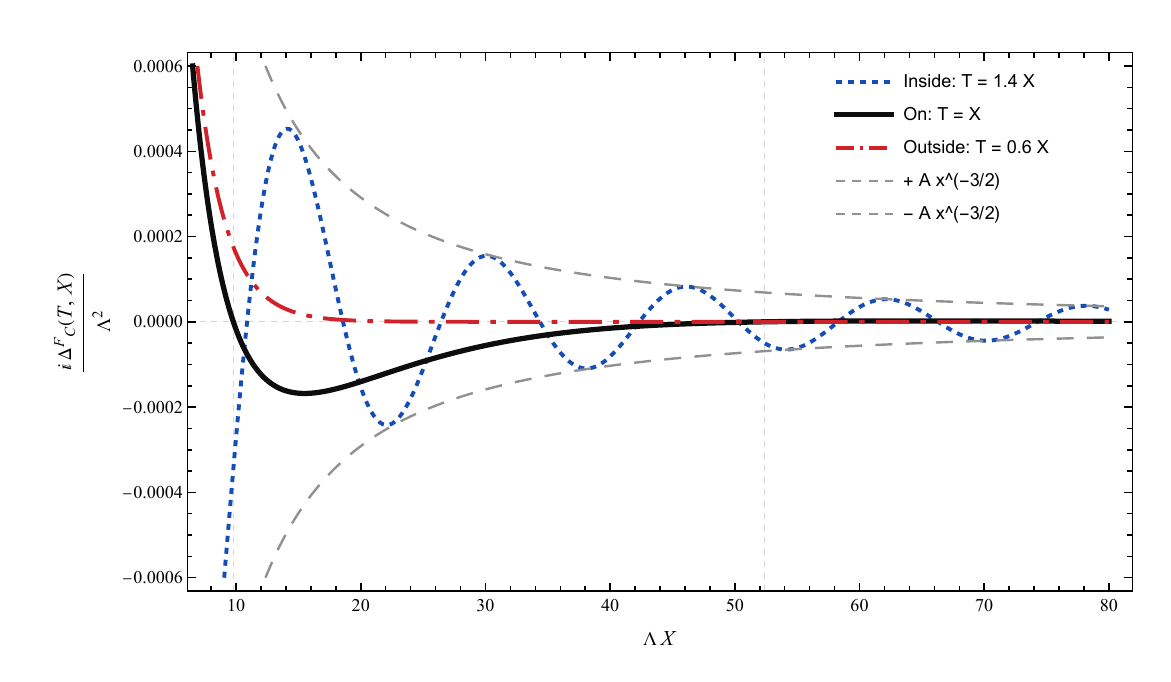}

\captionof{figure}{Signed Pauli--Jordan commutator for $D=3$, $n=2$, $r=6$, and $m/\Lambda=0.4$, evaluated along $T=(1+a)X$, $T=X$, and $T=(1-a)X$, with $a=0.4$, corresponding to the inside, on-cone, and outside regions, respectively. The left panel shows the behavior at smaller separations, while the right panel extends the comparison to larger distances. The gray dashed curves represent the leading stationary-phase envelopes $\pm A(\Lambda X)^{-3/2}$ of the timelike contribution. Its oscillatory behavior contrasts with the exponential suppression of the on-cone and spacelike amplitudes.}
\label{fig:pj-massive-comparison}
\end{minipage}

\clearpage
\twocolumngrid

\subsection{The massless result in three spatial dimensions}\label{sec:massless}
For $m=0$, $D=3$ and $n=2$, the entire family admits a closed expression. Indeed, setting $\nu=r/2$ and
\begin{equation}\label{eq:A-r}
\begin{split}
    \mathcal A_r(z)&=\int_0^\infty\dd k\,
    \frac{\cos(kz)}{(1+k^2/\Lambda^2)^{r/2}}\\
    &=\frac{\sqrt\pi\,\Lambda}{\Gamma(\nu)}
    \left(\frac{\Lambda|z|}{2}\right)^{\nu-1/2}
    K_{\nu-1/2}(\Lambda|z|),
\end{split}
\end{equation}
we find, for $T\ge0$ and $X>0$,
\begin{equation}\label{eq:massless-exact}
    \Delta_C^F(T,X)=-\frac{\ii}{4\pi^2X}
    \left[\mathcal A_r(X-T)-\mathcal A_r(X+T)\right].
\end{equation}
For $T<0$ the result is determined by oddness in $T$. At the Snyder value $r=4$,
\begin{equation}\label{eq:A4}
    \mathcal A_4(z)=\frac{\pi\Lambda}{4}
    (1+\Lambda|z|)\mathrm e^{-\Lambda|z|}.
\end{equation}
Thus the massless commutator depends essentially on the two light-cone combinations $X\pm T$. Outside the cone, for $\delta=X-T>0$, $\Lambda\delta\gg1$, and $\Lambda T\gg1$ (so the second term is exponentially smaller), its leading behavior is
\begin{equation}\label{eq:massless-yukawa}
    \Delta_C^F\sim-
    \frac{\ii\Lambda}{4\pi\,2^{r/2}\Gamma(r/2)X}
    (\Lambda\delta)^{r/2-1}\mathrm e^{-\Lambda\delta},
    \quad \Lambda\delta\gg1.
\end{equation}
This is the desired Yukawa-like spacelike suppression, valid for any $r>1$ under the stated asymptotic conditions; the infrared and ultraviolet convergence conditions must also be imposed when evaluating the full field theory. Notice that the exponential depends on the \emph{distance outside the light cone}, not simply on $X$ when $T$ increases with $X$.

There is also a precise statement about the hierarchy in the massless sector. For equal inside and outside displacements $a>0$ and $X>a$, the expressions at $T=X+a$ and $T=X-a$ differ only in the terms $\mathcal A_r(2X+a)$ and $\mathcal A_r(2X-a)$. Since $\mathcal A_r(z)$ is positive and strictly decreasing for $z>0$, the inside amplitude is strictly larger than the outside one. Furthermore, at sufficiently large $\Lambda X$, with $a$ fixed, $\mathcal A_r(0)>\mathcal A_r(a)$ while both distant-image terms are exponentially small. We therefore obtain
\begin{equation}\label{eq:massless-hierarchy}
    |\Delta_C^F|_{\rm on}>
    |\Delta_C^F|_{\rm inside}>
    |\Delta_C^F|_{\rm outside},
\end{equation}
in this asymptotic near-cone regime. The on-cone amplitude need not dominate at more modest separations, as illustrated by the numerical results below.

A similar comparison can be made by considering $T=(1\pm a)X$, now with $0<a<1$ dimensionless. For small $X$, the commutator behaves as $\Delta_C^F(T,X)\simeq-\ii T\mathcal K_F(0)$, provided $r>D$, so that the inside amplitude initially dominates. At large separation, however, the on-cone contribution decreases as $X^{-1}$, whereas the inside and outside amplitudes become exponentially suppressed. Consequently, the on-cone and inside curves must cross, while the inside amplitude remains greater than the outside one. This makes the change of hierarchy a natural consequence of the different trajectories relative to the light cone, rather than a peculiarity of the numerical results.

\begin{figure*}[t]
\centering
\includegraphics[
        width=0.49\textwidth,
        trim=1cm 0cm 0.5cm 0cm,
        clip
    ]
    {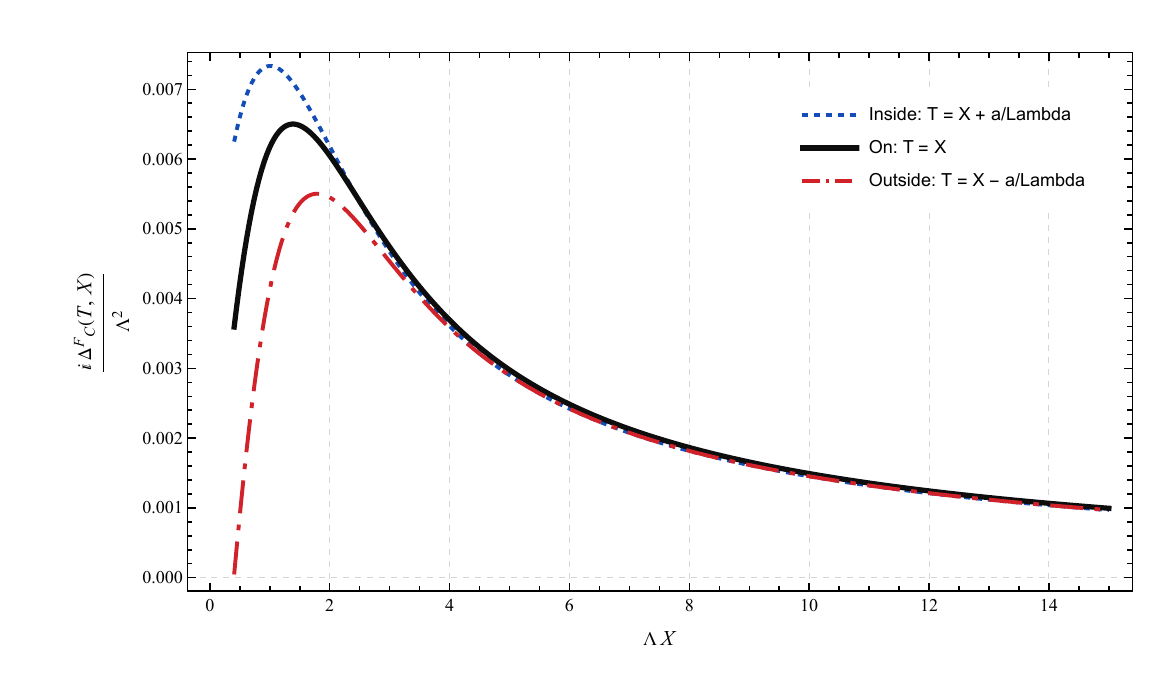}
\hfill
\includegraphics[
        width=0.49\textwidth,
        trim=1cm 0cm 0.5cm 0cm,
        clip
    ]
    {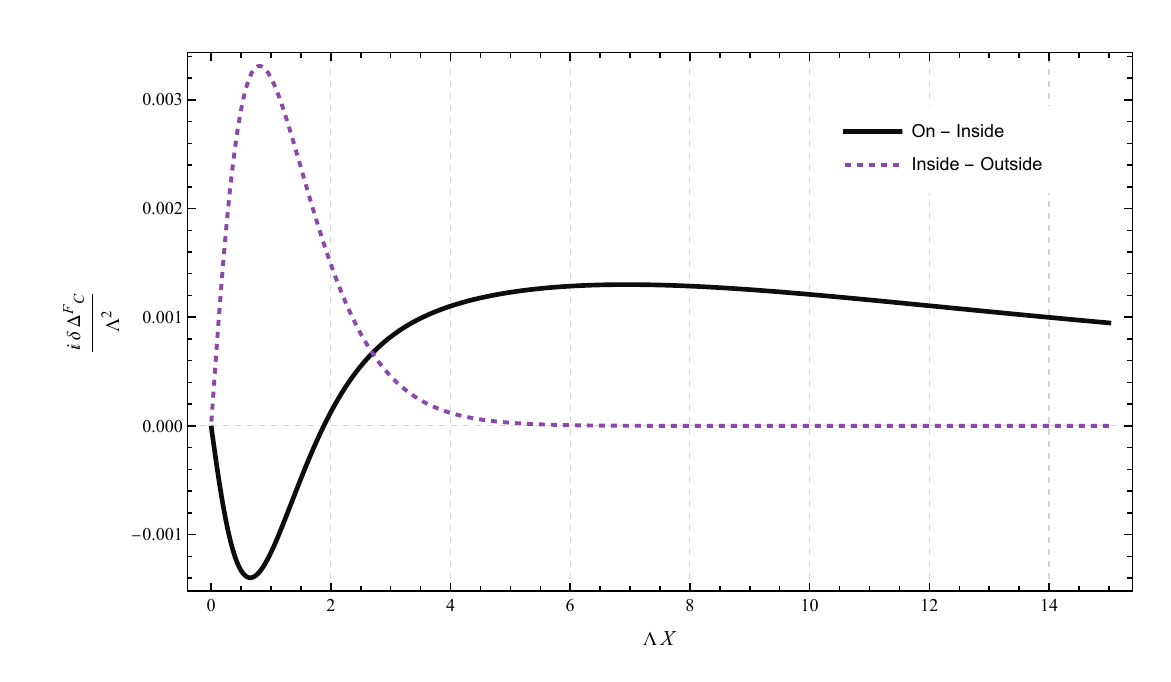}
\caption{Massless Pauli--Jordan commutator for $D=3$, $n=2$, and $r=6$. The left panel compares the signed amplitudes inside, on, and outside the light cone, using $T=X+a/\Lambda$, $T=X$, and $T=X-a/\Lambda$, respectively, with $a=0.4$. The right panel shows the differences between the on-cone and inside amplitudes (solid curve), and between the inside and outside amplitudes (dashed curve), now taking $T=(1\pm a)X$. The latter difference remains positive, whereas the former changes sign, reflecting the transition between the short-distance and asymptotic hierarchies.}
\label{fig:pj-massless-comparison}
\end{figure*}

The change in the massless hierarchy is illustrated in Fig.~\ref{fig:pj-massless-comparison}. The fixed-displacement comparison shows the relative amplitudes near the light cone, while the second panel considers trajectories with a constant ratio $T/X$. In the latter case, the difference between the on-cone and inside amplitudes changes sign, whereas the inside amplitude remains greater than the outside one, in agreement with the preceding analytical discussion.

\section{An illustrative application: self-interacting fields}\label{sec:applications}
The preceding construction can be extended beyond the free theory. In particular, the observer-covariant interpretation allows us to examine the ultraviolet behavior of a simple interaction without introducing a modified free dispersion relation. Let $\psi=\phi_{F,u}$ and consider
\begin{equation}\label{eq:interaction}
    S[\psi;u]=S_0[\psi;u]-\frac g2
    \int\dd^{D+1}x\,(\psi^\dagger\psi)^2,
    \qquad g\ge0,
\end{equation}
where $S_0[\psi;u]$ is the covariant extension of the quadratic action \eqref{eq:filtered-action}. For a constant $u^\mu$, it may be written as
\begin{equation}\label{eq:observer-action}
    S_0[\psi;u]=\int\dd^{D+1}x\,\bigl[
    (\partial_\mu\psi^\dagger)
    \mathcal T_{\Lambda,u}^{-2}(\partial^\mu\psi)
    -m^2\psi^\dagger\mathcal T_{\Lambda,u}^{-2}\psi\bigr].
\end{equation}
Under simultaneous Lorentz transformations of the coordinates, $u^\mu$, and the scalar field $\psi$, the complete action is invariant. This observer covariance should not be confused with invariance under the nonlocal field transformation obtained by holding $u^\mu$ fixed and conjugating ordinary boosts with $\mathcal T_{\Lambda,u}$; the quartic term is not generally invariant under that transformation.

In the rest frame $u^\mu=(1,\mathbf0)$, the free propagator is Eq.~\eqref{eq:G-momentum}. Equivalently, writing $\psi=\mathcal T_{\Lambda,u}\phi$ gives the ordinary Klein--Gordon free action for the auxiliary field $\phi$, together with a nonlocal quartic interaction. Each external momentum entering the corresponding vertex carries one factor $\sqrt{F_\Lambda(q(k,u))}$. Thus the interaction is not merely the ordinary local $\phi^4$ theory rewritten in different variables.

As a first application, the one-loop tadpole integral is proportional to
\begin{equation}\label{eq:tadpole}
    I_{\rm tad}(m)=\int\frac{\dd^Dk}{(2\pi)^D}
    \frac{F_\Lambda(k)}{2\omega_k}
    =\langle\phi_F\phi_F^\dagger\rangle,
\end{equation}
up to the coupling and the combinatorial factor determined by the normalization of the interaction. It is finite for $r>D-1$, and its explicit expressions are those obtained in Sec.~\ref{sec:vacuum}. The stronger condition $r>D+1$ also makes the weighted free vacuum-energy density finite. The tadpole is independent of the inertial coordinates used to evaluate it when $u^\mu$ is transformed along with the momenta. Indeed, the corresponding off-shell free propagator is $G_{F,u}(p)=\ii F_\Lambda(q(p,u))/(p_\mu p^\mu-m^2+\ii\epsilon)$, and the weight is a scalar under these simultaneous transformations.

If the quartic term is regarded as a fundamental interaction, the vector $u^\mu$ is additional physical structure and its interpretation must be specified; a detector-dependent filter should not automatically be identified with a universal modification of particle dynamics. This result does not establish ultraviolet finiteness of the complete interacting theory, nor does observer covariance alone provide a physically distinct DSR realization. In particular, counterterms, overlapping divergences, unitarity, and the causal structure of interacting observables remain to be examined. Nevertheless, the tadpole supplies a first example in which the same spectral function governing the free-field fluctuations also suppresses a radiative contribution. The extension to spinor and gauge fields, together with their corresponding identities, will be considered elsewhere.

\section{Conclusions and perspectives}\label{sec:conclusions}
We have reconsidered the weighted-measure formulation of quantum field theory by introducing a dimensionless function with an independent momentum scale. In contrast with the constant-power constructions of Refs.~\cite{Cartas2022,Cartas2025}, the present family restores the standard spectral measure at low momenta and separates the ultraviolet exponent from the scale at which the deformation becomes important.

The main results can be summarized as follows. First, the quadratic weight admits a direct algebraic motivation: for $r=D+1$ it agrees with the momentum-space measure for a standard Euclidean Snyder representation, while $r>D+1$ ensures ultraviolet convergence of the weighted vacuum energy. Second, the connection between measure deformations, modified dispersion relations, and nonlinear momenta is made explicit. An observer-covariant extension of the same weight is obtained from the spatial momentum measured by a timelike observer. This preserves the standard mass shell and transforms covariantly when the observer four-velocity is transformed together with the field, although it does not by itself define a physically distinct DSR theory. Third, we obtain finite spatial commutator kernels, vacuum field fluctuations, and weighted vacuum energies, with their precise convergence conditions in arbitrary spatial dimension.

The two-point sector further reveals a natural spatial-resolution interpretation. The propagator is a spatial convolution of its ordinary counterpart, and the Pauli--Jordan commutator exhibits an exponentially suppressed, but generally nonvanishing, spacelike tail. For $n=2$ the massless result is available in closed form for arbitrary $r$ in three spatial dimensions, and the massive spacelike amplitude is explicitly reducible to derivatives of the Yukawa kernel for even $r$. Therefore the construction softens the short-distance singular structure at the cost of exact microcausality in the filtered observable algebra.

The distinction between the filtered energy and the genuine generator of time translations is also essential. It prevents identifying the finite expectation value of a spectrally resolved observable with a proof that the complete canonical vacuum-energy problem has disappeared. In the same spirit, the observer-covariant interacting example demonstrates ultraviolet improvement at the level of a simple one-loop contribution but leaves open the question of renormalization to all orders, as well as the physical status of the additional observer four-velocity.

Several directions remain for future work. The relation to a fully consistent DSR dynamics could be strengthened by fixing the physical energy--momentum variables and their nonlinear composition law. Other interesting extensions include the construction of a compatible noncommutative phase space, the analysis of nonzero particle--antiparticle asymmetry, the quantization of spinor and vector fields, and interacting theories with the corresponding symmetries. A more detailed numerical investigation of the light-cone hierarchy for different $r$ and $m/\Lambda$ would also be worthwhile. These questions will help determine to what extent the weighted measure can be interpreted as more than a phenomenological finite-resolution prescription.

\onecolumngrid
\appendix
\section{Symmetry of the Snyder representation}\label{app:snyder}
For completeness, write $A_{ij}=\delta_{ij}+k_i k_j/\Lambda^2$ and consider the operator $x_i=\ii A_{ij}\partial_{k_j}$. Assuming boundary terms vanish, its symmetry in $L^2(\dd^Dk\,w)$ requires
\begin{equation}\label{eq:app-snyder-condition}
\partial_{k_j}(w A_{ij})=0.
\end{equation}
For an isotropic function $w=w(k^2)$ one has
\begin{align}
    \partial_{k_j}(w A_{ij})
    &=\partial_{k_i}w+\frac1{\Lambda^2}\left[k_i k_j\partial_{k_j}w+(D+1)k_i w\right]\\
    &=k_i\left[2\left(1+\frac{k^2}{\Lambda^2}\right)\frac{\dd w}{\dd(k^2)}+\frac{D+1}{\Lambda^2}w\right] =0.
\end{align}
Thus $\dd\ln w/\dd(k^2)=-(D+1)/[2(\Lambda^2+k^2)]$, and integration gives $w=C(1+k^2/\Lambda^2)^{-(D+1)/2}$. With the normalization $w(0)=1$, $C=1$. This establishes Eq.~\eqref{eq:snyder-measure} for arbitrary $D$.

The preceding construction can be extended to the more general family of measures considered in the main text. Let
\begin{equation}
    w_r(k)=
    \left(1+\frac{k^2}{\Lambda^2}\right)^{-r/2},
\end{equation}
a direct calculation gives
\begin{equation}
    \partial_{k_j}(w_r A_{ij})
    =\frac{D+1-r}{\Lambda^2}k_iw_r.
\end{equation}
Consequently, the operator
\begin{equation}\label{eq:snyder-general}
    \hat x_i^{(r)}=\ii\left[
    A_{ij}\partial_{k_j}
    +\frac{D+1-r}{2\Lambda^2}k_i
    \right]
\end{equation}
is symmetric with respect to the scalar product
\begin{equation}
    \langle f,g\rangle_r=
    \int_{\mathbb R^D}\dd^Dk\,
    w_r(k)f^*(\mathbf k)g(\mathbf k),
\end{equation}
provided the corresponding boundary terms vanish.

The additional multiplication term does not modify the Snyder commutation relations. In particular,
\begin{equation}
    [\hat x_i^{(r)},\hat k_j]
    =\ii\left(\delta_{ij}
    +\frac{\hat k_i\hat k_j}{\Lambda^2}\right),
\end{equation}
while the original representation is recovered for $r=D+1$.

In fact, these representations are unitarily related. Writing $r_0=D+1$, the transformation
\begin{equation}
    V_r=
    \left(\frac{w_{r_0}}{w_r}\right)^{1/2}
    =\left(1+\frac{k^2}{\Lambda^2}\right)^{
    (r-D-1)/4}
\end{equation}
maps the Hilbert space with weight $w_{r_0}$ onto that with weight $w_r$, and satisfies
\begin{equation}
    \hat x_i^{(r)}
    =V_r\hat x_i^{(r_0)}V_r^{-1}.
\end{equation}
Thus, varying $r$ preserves the Snyder algebra and corresponds to a change of representation, rather than a new deformation of its commutation relations.

\section{Fourier kernels and convergence integrals}\label{app:integrals}
For $n=2$, the Gamma representation
\begin{equation}
    \left(1+\frac{k^2}{\Lambda^2}\right)^{-r/2}
    =\frac1{\Gamma(r/2)}\int_0^\infty\dd u\,u^{r/2-1}
    \mathrm e^{-u(1+k^2/\Lambda^2)}
\end{equation}
reduces the Fourier transform to a Gaussian integral,
\begin{equation}
    \wt(X)=\frac{\Lambda^D}{(4\pi)^{D/2}\Gamma(r/2)}
    \int_0^\infty\dd u\,u^{(r-D)/2-1}
    \mathrm e^{-u-(\Lambda X)^2/(4u)}.
\end{equation}
Applying the integral representation of $K_\nu$ gives Eqs.~\eqref{eq:kernel-bessel} and \eqref{eq:kernel-zero}. For $r=D+1$, the identity $K_{1/2}(z)=\sqrt{\pi/(2z)}\mathrm e^{-z}$ yields Eq.~\eqref{eq:snyder-kernel}.

The general massless moment formula is
\begin{equation}
    \int_0^\infty\dd k\,k^{a-1}F_\Lambda(k;n,r)
    =\frac{\Lambda^a}{n}B\left(\frac an,\frac{r-a}n\right),\quad 0<a<r,
\end{equation}
from the substitution $z=(k/\Lambda)^n/[1+(k/\Lambda)^n]$. This gives Eqs.~\eqref{eq:VEV-zero-m} and \eqref{eq:rho-zero-m} immediately.

For $n=2$ and $m>0$, set $z=k^2/(k^2+m^2)$ and $\xi=1-m^2/\Lambda^2$. The general integral
\begin{equation}
    \int_0^\infty\dd k\,k^{D-1}(k^2+m^2)^{\sigma/2}
    \left(1+\frac{k^2}{\Lambda^2}\right)^{-r/2}
\end{equation}
becomes
\begin{equation}\label{eq:app-general-integral}
    \frac{m^{D+\sigma}}2 B\left(\frac D2,\frac{r-D-\sigma}  {2}\right)
    {}_2F_1\left(\frac r2,\frac D2;\frac{r-\sigma}{2};\xi\right),
    \qquad r>D+\sigma.
\end{equation}
Taking $\sigma=-1$ and $\sigma=+1$ supplies, respectively, the massive vacuum field fluctuation and the massive weighted energy density after multiplication by their prefactors in Eqs.~\eqref{eq:VEV-general} and \eqref{eq:rho-general}. The expressions are understood by continuity when $m=\Lambda$.

\section{Equal-time commutators and asymmetric two-point functions}\label{app:algebra}
Write $u_{\mathbf k}(x)=\mathrm e^{-\ii\omega_kt+\ii\mathbf k\cdot\mathbf x}/\sqrt{2\omega_k}$ and use independent creation--annihilation algebras. For the asymmetric expansion
\begin{equation}
\phi_F=\int\frac{\dd^Dk}{(2\pi)^{D/2}}\left[\sqrt{F_+(k)}\,a_{\mathbf k}u_{\mathbf k}+
\sqrt{F_-(k)}\,b_{\mathbf k}^\dagger u_{\mathbf k}^*\right],
\end{equation}
we obtain at equal time ($\mathbf X=\mathbf x-\mathbf y$)
\begin{align}
[\phi_F(t,\mathbf x),\dot\phi_F^\dagger(t,\mathbf y)]
&=\frac{\ii}{2}\int\frac{\dd^Dk}{(2\pi)^D}
\left[F_+\mathrm e^{\ii\mathbf k\cdot\mathbf X}+F_-\mathrm e^{-\ii\mathbf k\cdot\mathbf X}\right]
=\ii\wt(\mathbf X),\\
[\phi_F(t,\mathbf x),\phi_F^\dagger(t,\mathbf y)]
&=\frac12\int\frac{\dd^Dk}{(2\pi)^D\omega_k}
\left[F_+\mathrm e^{\ii\mathbf k\cdot\mathbf X}-F_-\mathrm e^{-\ii\mathbf k\cdot\mathbf X}\right]
=\mathcal K_{\rm NC}^{\phi}(\mathbf X),\\
[\dot\phi_F(t,\mathbf x),\dot\phi_F^\dagger(t,\mathbf y)]
&=\frac12\int\frac{\dd^Dk}{(2\pi)^D}\omega_k
\left[F_+\mathrm e^{\ii\mathbf k\cdot\mathbf X}-F_-\mathrm e^{-\ii\mathbf k\cdot\mathbf X}\right]
=\mathcal K_{\rm NC}^{\pi}(\mathbf X).
\end{align}
Here we used evenness in $\mathbf k$ together with $F_\pm=F_\Lambda(1\pm\eta)$. For arbitrary time separations the Hadamard and commutator functions are
\begin{align}
\Delta_H^F(T,\mathbf X)&=\int\frac{\dd^Dk}{(2\pi)^D\omega_k}F_\Lambda(k)
\left[\cos(\omega_kT-\mathbf k\cdot\mathbf X)-\ii\eta(k)\sin(\omega_kT-\mathbf k\cdot\mathbf X)\right],\\
\Delta_C^F(T,\mathbf X)&=\int\frac{\dd^Dk}{(2\pi)^D\omega_k}F_\Lambda(k)
\left[\eta(k)\cos(\omega_kT-\mathbf k\cdot\mathbf X)-\ii\sin(\omega_kT-\mathbf k\cdot\mathbf X)\right].
\end{align}
For $\eta=0$ the second equation reduces to Eq.~\eqref{eq:pj-general}. In particular, $\Delta_C^F(0,\mathbf X)=0$ for the symmetric theory, without implying spacelike microcausality.

\section{Retarded identity and analytic massless tails}\label{app:causal}
Let $g_\omega(T)=\sin(\omega T)/\omega$ for $T\ge0$. Its Laplace transform is $(z^2+\omega^2)^{-1}$. Because $\mathcal L\{S_\Lambda\}=(z^2+m^2-\Lambda^2)^{-1}$, the identity
\begin{equation}
\frac{1}{(k^2+\Lambda^2)(z^2+k^2+m^2)}
=\frac1{z^2+m^2-\Lambda^2}
\left[\frac1{k^2+\Lambda^2}-\frac1{z^2+k^2+m^2}\right]
\end{equation}
implies, on inverse Laplace transformation,
\begin{equation}
\frac{g_{\omega_k}(T)}{k^2+\Lambda^2}
=\frac{S_\Lambda(T)}{k^2+\Lambda^2}
-\int_0^T\dd s\,S_\Lambda(s)g_{\omega_k}(T-s),\qquad T>0.
\end{equation}
Taking the spatial Fourier transform and restoring oddness under $T\mapsto-T$ produces Eq.~\eqref{eq:pj-resolvent} with the retarded Green function normalized by $G_D^{\rm ret}(T,\mathbf X)=\theta(T)\int\dd^Dk\,(2\pi)^{-D}\,g_{\omega_k}(T)\mathrm e^{\ii\mathbf k\cdot\mathbf X}$. This derivation also shows why the $\operatorname{sgn}(T)$ multiplies the convolution term when the upper integration endpoint is $|T|$.

For $m=0$, $D=3$, Eq.~\eqref{eq:pj-general} gives
\begin{equation}
    \Delta_C^F(T,X)=-\frac{\ii}{2\pi^2X}\int_0^\infty\dd k\,
    F_\Lambda(k)\sin(kX)\sin(kT).
\end{equation}
Applying $\sin(kX)\sin(kT)=[\cos k(X-T)-\cos k(X+T)]/2$ proves Eq.~\eqref{eq:massless-exact}. Finally the large-$z$ asymptotic $K_\alpha(z)\sim\sqrt{\pi/(2z)}\mathrm e^{-z}$ yields
\begin{equation}
    \mathcal A_r(z)\sim\frac{\pi\Lambda}{2^{r/2}\Gamma(r/2)}
    (\Lambda|z|)^{r/2-1}\mathrm e^{-\Lambda|z|},
    \qquad \Lambda|z|\gg1,
\end{equation}
which is the basis of the explicit Yukawa-like tail \eqref{eq:massless-yukawa}.

\section*{Acknowledgments}
R. C. F. acknowledges support from the Sistema Nacional de Investigadores (Mexico). S. G. S. would like to thank VIEP-BUAP (Mexico) for financial support. Numerical computations and figures were produced using Mathematica.

The authors acknowledge the use of ChatGPT (OpenAI) for assistance with linguistic revision and the exploration of certain mathematical ideas, including the application of the Bochner--Schoenberg theorem to the positive definiteness of the kernel and the evaluation of analytical integrals. All AI-generated suggestions, derivations, and results were independently examined and verified by the authors, who take full responsibility for the content of this work.

\bibliography{references}

@article{Cartas2022,
  author        = {Cartas-Fuentevilla, R. and Mendez-Ugalde, A.},
  title         = {Lorentz-breaking weighted measures as quantum field theory regulators},
  journal       = {Phys. Lett. B},
  volume        = {830},
  pages         = {137146},
  year          = {2022},
  doi           = {10.1016/j.physletb.2022.137146},
  eprint        = {2201.11588},
  archivePrefix = {arXiv}
}

@article{Cartas2025,
  author        = {Cartas-Fuentevilla, R. and Gonz{\'a}lez-Salud, S. and
                   B{\'a}rcena-Ramos, R. and Berra-Montiel, J.},
  title         = {The propagator field theory revisited: A {Lorentz} symmetry breaking approach},
  journal       = {Ann. Phys. (N.Y.)},
  volume        = {479},
  pages         = {170075},
  year          = {2025},
  doi           = {10.1016/j.aop.2025.170075},
  eprint        = {2409.10313},
  archivePrefix = {arXiv}
}

@article{Snyder1947,
  author  = {Snyder, H. S.},
  title   = {Quantized space-time},
  journal = {Phys. Rev.},
  volume  = {71},
  pages   = {38--41},
  year    = {1947},
  doi     = {10.1103/PhysRev.71.38}
}

@article{LuStern2012,
  author        = {Lu, Lei and Stern, A.},
  title         = {Particle dynamics on {Snyder} space},
  journal       = {Nucl. Phys. B},
  volume        = {860},
  pages         = {186--205},
  year          = {2012},
  doi           = {10.1016/j.nuclphysb.2012.02.012},
  eprint        = {1110.4112},
  archivePrefix = {arXiv}
}

@article{MagueijoSmolin2002,
  author        = {Magueijo, J. and Smolin, L.},
  title         = {{Lorentz} invariance with an invariant energy scale},
  journal       = {Phys. Rev. Lett.},
  volume        = {88},
  pages         = {190403},
  year          = {2002},
  doi           = {10.1103/PhysRevLett.88.190403},
  eprint        = {hep-th/0112090},
  archivePrefix = {arXiv}
}

@article{MeasureMatters,
  author        = {Amelino-Camelia, G. and Arzano, M. and
                   Gubitosi, G. and Magueijo, J.},
  title         = {Dimensional reduction in momentum space and scale-invariant cosmological fluctuations},
  journal       = {Phys. Rev. D},
  volume        = {88},
  pages         = {103524},
  year          = {2013},
  doi           = {10.1103/PhysRevD.88.103524},
  eprint        = {1309.3999},
  archivePrefix = {arXiv}
}

@article{DimReduction,
  author        = {Amelino-Camelia, G. and Arzano, M. and
                   Gubitosi, G. and Magueijo, J.},
  title         = {Dimensional reduction in the sky},
  journal       = {Phys. Rev. D},
  volume        = {87},
  pages         = {123532},
  year          = {2013},
  doi           = {10.1103/PhysRevD.87.123532},
  eprint        = {1305.3153},
  archivePrefix = {arXiv}
}

@article{Kempf1995,
  author        = {Kempf, A. and Mangano, G. and Mann, R. B.},
  title         = {Hilbert space representation of the minimal length uncertainty relation},
  journal       = {Phys. Rev. D},
  volume        = {52},
  pages         = {1108--1118},
  year          = {1995},
  doi           = {10.1103/PhysRevD.52.1108},
  eprint        = {hep-th/9412167},
  archivePrefix = {arXiv}
}

@article{Schoenberg1938,
  author  = {Schoenberg, I. J.},
  title   = {Metric spaces and positive definite functions},
  journal = {Trans. Am. Math. Soc.},
  volume  = {44},
  pages   = {522--536},
  year    = {1938},
  doi     = {10.1090/S0002-9947-1938-1501980-0}
}

@article{ColladayKostelecky1998,
  author        = {Colladay, D. and Kosteleck{\'y}, V. A.},
  title         = {{Lorentz}-violating extension of the standard model},
  journal       = {Phys. Rev. D},
  volume        = {58},
  pages         = {116002},
  year          = {1998},
  doi           = {10.1103/PhysRevD.58.116002},
  eprint        = {hep-ph/9809521},
  archivePrefix = {arXiv}
}
\end{document}